\documentclass[12pt]{article}
\title{Comments on Dumitrescu's ``A Selectable Sloppy Heap''}

\author{
Michael L. Fredman\\
Department of Computer Science\\
Rutgers University, New Brunswick\\
\small Email: fredman@cs.rutgers.edu}

\begin{document}
%\title{Comments on Dumitrescu's ``A Selectable Sloppy Heap''}
%\author{ Michael L. Fredman}
%\institute{Rutgers University, New Brunswick\\ \email{fredman@cs.rutgers.edu}}
\maketitle

\begin{abstract}
Dumitrescu [arXiv:1607.07673] describes a data structure referred to as a {\em Selectable
Sloppy Heap\/}.  We present a simplified approach, and also point out
aspects of Dumitrescu's exposition that require scrutiny.
\end{abstract}

{\bf Introduction}

Given a constant $k >1$, Dumitrescu [arXiv:1607.07673] defines 
(modifying his terminology) a $k$-Selectable Sloppy heap 
to be a data structure that maintains a dynamic 
set $S$ of items having keys belonging to an ordered universe,
that supports the operations Delete-$i$ ($1 \leq i \leq k$), and 
insertion.  The Delete-$i$ operation returns and deletes from $S$
an item whose key belongs to its $i$-th quantile: partitioning $S$ into
$k$ uniformly sized intervals, the $i$-th quantile consists of the 
$i$-th such interval (from the left). The data structure 
implementation determines the particular
choice of the item selected for removal when performing a Delete-$i$
operation.  

We see in Dumitrescu's formulation, which extends a homework problem
devised by this author (Fredman), an aesthetically attractive topic.
Dumitrescu's posting concerns the design of a data structure that
supports these operations with 
worst-case $O(\log k)$ costs, independently of $n$. 
A simple argument, left for the reader, shows that the
there is an inherent $\Omega (\log k)$ amortized cost 
per operation for worst-case sequences of these
operations in the comparison-based model of computation, 
demonstrated by reduction from sorting.

There are issues, however, with the posting which we proceed to 
describe.  First, a brief overview. As described 
in greater detail below, the data structure
design is centered on a balanced binary tree where the intent is to 
maintain $O(k)$ buckets consisting of item intervals at the 
leaves; a requested
operation accesses, at cost $O(\log k)$, the appropriate bucket 
to insert or delete, 
as the case may be, 
an item belonging to the bucket. (No ordering is maintained within
single buckets.)
Bucket splitting takes place when necessary so that the buckets remain
sufficiently small, guaranteeing that at least one falls within
any given quantile, thereby correctly 
supporting deletion requests.  To maintain
the number of buckets at $O(k)$, merging of small buckets also takes
place.  The work required of these costly merging and splitting tasks 
is distributed over operation requests so as to reduce the worst-case 
costs to same magnitude as that of the attainable amortized costs.  

Dumitrescu's method implements a rapidly moving round-robin 
process among
the buckets (transitioning one bucket with every requested operation)
for the purpose of detecting buckets in need of splitting.  While it 
would seem natural to detect adjacent small buckets appropriate for
merging as they are encountered within the same round-robin processing, 
this is not done.
Instead a secondary structure, a priority queue, is maintained 
that stores the combined sizes of adjacent
bucket pairs, and with each requested operation the smallest such
sum in the priority queue is checked for
falling below a specified threshold, and if so, 
the corresponding pair of buckets gets merged.  It is in
consideration of this priority queue that two errors appear
in the posting.  First, the posting only requires updates to the 
priority queue in the context of bucket merges and splits, but updates
are necessary with {\em every\/} requested operation, since each changes
the size of some bucket. This constant need of updating the priority
queue by itself entails as much work as the tree searching costs. 
A second issue concerns the proof that the number of buckets
is $O(k)$. That proof proceeds with an induction step that relies upon
just one bucket being split over the course of 
a single operation request.  However,
two buckets can potentially get split; both the round-robin bucket
{\em and\/} the bucket at which a requested operation takes place
(both undergoing processing with each requested operation).  

Fortunately, there is no need for maintaining the secondary 
priority queue structure involving bucket sizes.  As indicated in
Dumitrescu's posting, this author (Fredman) had in mind a different
strategy that performed, in essence, operation-distributed periodic 
reorganizations of the data structure to restrain the proliferation
of small buckets, without utilizing explicit bucket merging.  Presented 
below is a modification of that method that {\em does\/} utilize bucket 
merging and therefore bears similarity to Dumitrescu's round-robin,
but proceeds at a more leisurely pace.  An advantage is that adjacent 
buckets that can be merged are detected as they are encountered, 
dispensing with a priority queue.  Dumitrescu's (apparent) 
purpose in utilizing a priority-queue 
is to facilitate an argument that bounds the number of buckets.  However
a simple potential-based argument accomplishes the same without having
to complicate the data structure.

\noindent
{\bf Our $O(\log k)$ worst-case Method}

So that this description is self-contained we review some basics, also
presented in Dumitrescu's posting [arXiv:1607.07673].

Our data structure stores its items in buckets positioned as the leaves 
of a balanced binary search tree. 
The items found in a given bucket
are consistent with the path in the tree leading to the bucket, but
order is not maintained within the buckets.  Each bucket stores
at most $n/(2k)$ items, so that for any requested Delete operation there
will be at least one bucket suitable for serving the request;
any of its items being suitable candidates for
deletion.  Each node in the binary 
tree also stores in a size field the number of items in the 
leaves of the subtree rooted
at that node. This field facilitates both insertions and deletions,
and also determination of the extent
of a merge operation (see below).  The buckets are also maintained 
in a doubly
linked list respecting their left-to-right positioning in the tree, so
that neighboring buckets with respect to tree order are immediately 
accessible from one-another.  When a bucket becomes empty its associated
leaf is deleted from the tree, and when a bucket splits a new leaf is
inserted into the tree. Finally, by making use of merging operations
the buckets will be maintained 
to be $O(k)$ in number, so that the required tree operations 
are all supported at $O(\log k)$ cost. 

This focus of this discussion centers on the
regime in which the number of items $n$ is bounded below by a 
(sufficiently) large multiple of $k$.
The conditions described above: namely that (a)
the maximum number of items in a bucket never exceeds $(1/2)n/k$, and
(b) the maximum number of buckets at any instant is $O(k)$;  
will be maintained
by scheduling supplemental work in the execution of requested 
operations.  This supplemental work may involve a fixed number of 
operations acting upon the binary tree, including insertion and
deletion of leaves (pointers to buckets), and joining and tree-splitting
operations.  These tree operations require work commensurate with that
of a tree search.  In addition to the tree operations, the supplemental
work includes a constant-bounded amount that acts upon and configures
buckets in isolation of the tree; 
namely splitting of large buckets and merging of small buckets. This
latter supplemental work is referred to as {\em bucket work\/}.
When a designated amount $w$ of such bucket work is to be performed, we
understand that this work is to be scheduled over some 
number of requested
operations.  When bounding that number of requested operations 
we will commonly encounter a term involving a positive multiplicative
constant that decreases inversely with the amount of bucket 
work that
gets scheduled per requested operation, and we use notation such as
$\lceil \epsilon \cdot w \rceil$ in specifying such a term, reflecting the 
understanding that $\epsilon$ can, by design, be made arbitrarily small.
(We omit the ceiling operator when the context of 
discussion clearly justifies doing so.)  The required number
of requested operations over which a given task gets accomplished
is referred to as its {\em requested-operation\/} cost.  Upon having
derived a bound, e.g. $O(\epsilon w)$, for requested operation cost,
with this interpretation for $\epsilon$ we are justified in abusing
formalism, restating the bound to be $\epsilon w$.
We will
likewise employ a multiplicative $\epsilon$ term when bounding the
amounts of other measures that are similarly subject to reduction.

{\bf Bucket Control} 

Overview: Bucket control is an uninterrupted process that 
proceeds in rounds, each referred to as a {\em
bucket control round\/}.  During one such round the buckets are
scanned from left to right, while bucket splitting and 
merging tasks are performed .  Other bucket splittings that interrupt 
the scanning
sequence can also take place.  The details of bucket control are 
discussed below.  The splitting and merging tasks are considered
first.

Bucket Splitting: When the size of a bucket reaches a defined 
threshold, it gets
split into two smaller buckets.  Depending on circumstances (discussed
below) the splitting process is distributed 
over a mix of some number of requested operations that directly 
access $B$, followed by some number of requested operations 
that don't necessarily access $B$.  Either number can be zero.
The sizes of the two buckets spawned by splitting are close in
size.  The splitting process is completed with tree operations
that replace the leaf pointing to the original bucket by
two consecutive leaves pointing to the respective spawned buckets.

Assume there are $m$ items in the bucket.  A fraction $\epsilon\dot m$ 
of the items are set aside to serve deletion requests while the 
splitting
takes place.  By design the splitting takes place at a sufficiently
rapid pace (large enough, but constant-bounded amount of supplemental work per 
requested operation) so that the set-aside items suffice to supply 
the 
deletion requests until the spawned buckets have been deployed.  
Insertions to the bucket are placed among the
set-aside items.  A linear-time median selection
algorithm is then applied to the remaining $(1-\epsilon)m$ items to 
obtain
a pivot value. This pivot is then used to partition the 
$m$ items, as well as items subsequently inserted as the
process proceeds.  So long as the rate at which bucket work takes
place is sufficiently fast, so that the pivot-based partitioning
repositions items faster than the rate at which they enter the 
bucket (at most one per requested operation),
we find that (a) the requested-operation cost for splitting a
bucket with $m$ items is bounded by $\lceil \epsilon m\rceil$; 
(b) the number of items in either spawned bucket does not exceed 
$((1/2)+\epsilon)m$;
and (c) prior to completion the number of items in the bucket at 
no point exceeds $(1 + \epsilon)m$.

Bucket Merging:  This task combines a collection of consecutive
small buckets into a single larger bucket.  
Generally the run of consecutive buckets being merged begins at some 
specified bucket $C$, and extends to the maximum number of
consecutive buckets
whose combined size does not exceed a defined {\em merging threshold\/}.
The first step applies joining and tree-splitting operations to remove 
the portion of the tree that spans the run of buckets, apart from $C$.
The size fields in the binary tree nodes facilitate the required
navigation to implement the fixed number 
tree-splitting and joining operations.  The items belonging
to any given bucket are stored in a linked list and $C$ is iteratively
grown, appending the constituent item lists of the removed buckets to 
the item list of $C$, which completes the merging process.
Apart from the fixed number of operations on the binary tree, the
additional bucket work to merge $j$ (say) buckets has a 
requested-operation cost bounded by $\lceil\epsilon j\rceil $. The
growing bucket $C$ serves access requests to the merged bucket while 
this post-merging bucket work is underway.

Bucket Control Round: A bucket control round scans the buckets,
splitting those that are too large, and merging consecutive buckets
that are too small.  The scan proceeds from left to right
passing through the existing buckets.  The quantity $\zeta = n'/(6k)$, 
where $n'$ is the number of items in the structure at the
onset of a given round, defines the targeted bucket size and
gets used as a parameter in setting the thresholds for splitting
and merging buckets.
A complete round has a requested operation cost bounded by
$\epsilon n'$ (demonstrated below), so that when taking the access 
effects of these requested operations into account, $n'$ changes
by a relatively small amount from one round to the next.

At the instant that the size of a bucket $B$ exceeds the 
{\em splitting threshold\/} $(5/3)\zeta$ it is designated for a splitting
procedure.  Thereafter the bucket work of every requested operation
that accesses $B$ is dedicated to its being split.  When the scanning
sequence of a round advances to a bucket $C$, then the bucket work
available from subsequent requested operations, when not preempted for
bucket splitting elsewhere, acts upon $C$ and its neighboring
buckets as follows.  If $C$ is designated as undergoing splitting, 
then this work completes the splitting, and the scanning advances to
the next bucket of the original list.
If starting at $C$ there is a run of consecutive buckets
whose combined size is at most the merging threshold (set to
$\zeta$), then the maximal such run
is merged and replaces $C$.  
Scanning then advances to the next
bucket that follows the merged run.  If neither splitting or 
merging is indicated, then scanning simply advances to the next bucket
beyond $C$.  

Analysis

Our analysis includes the claim, proven by induction, that at the 
end of a bucket control
round the number of items $n$ in the structure satisfies 
$|n-n'|/n' \leq 1/9$.
We define a potential $P$ as follows. 
$P = P_1 + P_2$, where $P_1$ is the sum of the bucket sizes 
at or to the right
of where the next scanning step is positioned to act; and $P_2$ is 
the sum, over all buckets, of the
{\em excesses\/} of the bucket sizes, where the excess size of a 
bucket $B$ is the positive amount, if any, that the size of $B$ exceeds
$(5/4)\zeta$. At the onset of a given round, with $n'$ items in the
structure and the scan positioned at the leftmost bucket,
the initial value for $P$ is clearly $O(n')$.
We argue first that any completed bucket splitting procedure
decreases $P$ by $\Omega (\zeta)$.  This follows by consideration of 
$P_2$, which has a term (corresponding to the bucket being split) 
initially set at $((5/3) -(5/4))\zeta$ immediately before the 
splitting gets underway (if the splitting was initiated in the current
round), that is then replaced 
by two terms having 0 contribution upon conclusion, since the sizes
of the spawned buckets are reduced, relative to the original,
essentially by a factor of 2. 
The access effects of the $\epsilon \zeta$ 
operations whose bucket work facilitates the splitting (reflecting 
the requested-operation cost of the
splitting task) may increase $P_2$ and also cause $P_1$ to increase,
but the total amount of increase is bounded by $\epsilon \zeta$. 
Thus, $P$ decreases by $\Omega(\zeta)$, as claimed.
If the bucket splitting work was initiated (but not completed) in the 
preceding round then the analysis needs to take into account the
splitting threshold applicable for that round.  Appealing to the
induction hypothesis, the relative change in $\zeta$, 
in comparison with its value
in the preceding round, is bounded by 1/9, so that in terms of its 
current value the applicable splitting threshold of the preceding
round lies in the interval $[(3/2)\zeta,(15/8)\zeta]$.  This does not
alter the conclusion that $P$ decreases by $\Omega(\zeta)$ when the
splitting gets completed in the current round.

Now upon considering any pair of consecutive scanning steps we find that
$P$ also decreases by $\Omega(\zeta)$.  To see this, observe that
at least one of the buckets left in the wake of the two 
scanning steps has size at least $(1/3)\zeta$; otherwise the items 
belonging to two of these buckets would have been combined into a single
bucket.  That particular bucket is responsible for an $\Omega(\zeta)$ 
decrease
in the $P_1$ term, and as before the access effects of the 
$\epsilon \zeta$ requested operations, whose combined bucket work
accomplishes the task required of the two scanning steps, 
don't alter this conclusion.  As for $P_2$, with respect
to merging buckets the terms in $P_2$ 
reflecting the directly involved buckets are all zero (before and after
the merging), and if neither
splitting nor merging take place, $P_2$ remains largely unchanged.

In addition to completed bucket splittings and scanning steps there
is the preempted bucket work for initiated, but not completed bucket
splitting procedures.  
For each such affected {\em swollen\/} bucket $B$
the requested-operation cost of this preempted bucket work performed
on $B$ (each such operation accessing $B$) is bounded
by $\epsilon\zeta$ and the 
contribution of $B$ to $P$ (i.e.\ $P_2$) upon completion of the 
round is $\Omega(\zeta)$.
Now during a given round, the $\Omega(\zeta)$ decreases to 
$P$ observed above for each
completed bucket splitting and each pair of scanning steps, and 
the residual
amount $\Omega(\zeta)$ attributable to each swollen bucket,
in conjunction with the $O(n')$ value of $P$ at its onset, 
imply that the total 
of three quantities: the number of 
scanning steps, the number of completed splitting procedures 
and the residual number of
swollen buckets;  is $O(k)$. This implies 
that the number of buckets at
the end of the round (and beginning of the next round) is $O(k)$; at
most one bucket per scanning step or completed 
splitting.  Moreover, the total requested-operation cost of
a given round is $\epsilon n'$ ($\epsilon \zeta$ per scanning step or
completed bucket splitting or swollen bucket).  The absolute difference 
between $n$, at the end of the round, and $n'$ can't
exceed this requested-operation cost, which by induction establishes
our claim that $|n-n'|/n' \leq 1/9$.
Provided that the initial round begins with $O(k)$ buckets, this 
assures that
at all times the number of buckets is $O(k)$. 
With the splitting threshold set at $(5/3)\zeta$ 
we can also ensure that the size of a bucket never 
exceeds $2\zeta < n/(2k)$ 
(including the case when
splitting is initiated in the preceding round);
sufficient for valid implementation of 
deletion operations.
\end{document}